\documentclass [aps,12pt,superscriptaddress,a4paper]{revtex4}
\usepackage{graphicx}  
\usepackage{color}  
\usepackage{subfigure}
\usepackage{float}
\usepackage{dcolumn}   
\usepackage{bm}        
\usepackage{amssymb}   
\usepackage{natbib}

\begin{document}
\title{Enhanced phase segregation induced by dipolar interactions in polymer blends}

\author{Rajeev Kumar}
\affiliation{Computer Science and Mathematics Division, Oak Ridge National Laboratory, Oak Ridge, TN-37831}
\email{kumarr@ornl.gov}
\affiliation{Center for Nanophase Materials Sciences, Oak Ridge National Laboratory, Oak Ridge, TN-37831}

\author{Bobby G. Sumpter}
\affiliation{Center for Nanophase Materials Sciences, Oak Ridge National Laboratory, Oak Ridge, TN-37831}

\author{M. Muthukumar}
\affiliation{Polymer Science and Engineering Department, University of Massachusetts, Amherst, MA-01002}

\date{\today}

\begin{abstract}

\noindent We present a generalized theory for studying phase separation in 
blends of polymers containing dipoles on their backbone. 
The theory is used to construct co-existence curves and to study the effects of dipolar interactions on 
interfacial tension for a planar interface between the coexisting phases. 
We show that a mismatch in monomeric dipole moments, or equivalently a mismatch in the dielectric constant of the pure components, leads to 
destabilization of the homogeneous phase. Corrections to the Flory-Huggins phase diagram are 
predicted using the theory. Furthermore, we show that the interfacial tension increases with 
an increase in the mismatch of the dipole moments of the components. Density profiles and interfacial tensions are constructed 
for diffuse and sharp polymer-polymer interfaces by extending the formalisms of Cahn-Hilliard and 
Helfand-Tagami-Sapse, respectively. 

\end{abstract}

\maketitle

\section{Introduction}
``Like dissolves like'' is one of the empirical rules\cite{intermolecular_forces} to determine 
the solubility of solutes in a solvent medium. 
Similarly, ``polar likes polar'' is the empirical rule\cite{intermolecular_forces} for determining 
the compatibility among different molecules. 
In this paper, we derive a theory of this rule in the context of polymers and provide a quantitative description of the 
role played by the polar nature of monomers in determining the commingling of polymers\cite{intermolecular_forces,
onsager_moments, dielectric}. The predictions of the theory are also relevant to the growing usage of dipolar polymers such as peptide-polymer conjugates\cite{peptide_poly_rev}, 
polypeptoids\cite{peptoid_rev}, and polyzwitterions\cite{polyzwitt_rev}.

When the polymer components do not bear any charges or dipoles, the phase behavior is predicted by the classical Flory-Huggins theory \cite{flory_book,scalingbook} by lumping all pair-wise short-ranged interactions into a set of `chi' parameters. This theory has been very successful in capturing various qualitative aspects of phase behavior of uncharged 
polymers, despite its well-known limitations. Although some improvements can be made by addressing composition 
fluctuations\cite{benoit_book,MM_1986} and monomer architecture\cite{KF_1990,KF_1991,KF_1998}, the presence of 
long-ranged forces due to charges and dipoles on the chain backbone makes the extension of 
the Flory-Huggins theory nontrivial. For polymeric systems with ionizable monomers, theories based on 
field-theoretic treatments\cite{MM_1996,warren,monica_2000,MM_2002,monica_2003,MM_2009,MM_2010}
have borne out to be successful in treating the phase behavior\cite{kaji_1988,delsanti_2000,prabhu_1,prabhu_2}. 
When the repeat groups of the polymer chains carry permanent dipoles, theoretical formulation becomes 
technically difficult primarily due to the anisotropy of the intersegment orientational 
interactions and due to its long-range nature. So far, only single chain of dipoles has been 
addressed\cite{MM_dipoles,podgornik_dipoles,MM_2004,kumar_glenn}. It was shown\cite{MM_dipoles} that the dipolar 
interactions can be treated as freely rotating at higher temperatures, but at lower temperatures chain 
orientations freeze into a multitude of frustrated conformations. 

In this paper, we consider only the high-temperature regime of rotating dipoles without being subjected to 
frustrated states and address the commingling of two polymer components bearing different dipole moments. We consider 
the inter-segment interaction to consist of two parts. The first part is the usual contribution from the excluded-volume effect (van der Waals interaction) responsible for the Flory-Huggins `chi' parameter, and the second part is due to point dipoles \cite{intermolecular_forces,orland_dipolar, 
kumar_kilbey}. The induced dipole moment arising from the polarizability of the molecules necessitating a self-consistent computation is ignored.  In addition to calculating the phase diagram, we have followed the formalisms of 
Cahn-Hilliard \cite{cahn_hilliard} and Helfand-Tagami-Sapse \cite{helfand_tagami1,helfand_tagami2,helfand_tagami3, helfand_sapse} and calculated the interfacial tension between two coexisting phases and the accompanying density profiles. We show that consideration of dipolar interactions
naturally leads to the introduction of
a local electric field dependent dielectric function\cite{onsager_moments,dielectric,orland_dipolar,kumar_kilbey,booth_1,booth_2,sandberg,fulton_approaches} in the theory along with
its dependence on the temperature and local number density of dipoles. Concentration 
dependence of the dielectric function is shown to have significant effects on the phase segregation 
and interfacial properties. In particular, we have shown that an increase in mismatch between the dipole moments, 
which is equivalent to an increase in dielectric mismatch between polymer pairs, leads to an enhanced phase segregation 
as well as sharper interfaces. As a result of sharper interfaces, interfacial tension for an interface between 
the coexisting phases increases with an increase in the dielectric mismatch. Also, these predictions are 
compared with theoretical treatment for freely rotating dipoles\cite{intermolecular_forces,MM_2004,MM_dipoles,kumar_glenn}, which is valid only at very high temperatures. 

The rest of the paper is organized as follows: the formalism is presented in 
section \ref{sec:theory}, followed by the results in section \ref{sec:results} and conclusions in section 
\ref{sec:conclusions}.

\section{Theory}\label{sec:theory}
\setcounter {equation} {0} 
We consider a polymer blend containing $n_A, n_B$ flexible chains of type $A$ and $B$, respectively. 
In this work, we consider monodisperse polymer chains so that each chain has $N_j$ Kuhn segments of 
length $b_j$ for $j= A,B$. 
Representing each chain as a continuous curve of length $N_j b_j$, an arc length variable $s_{j,\alpha}$
is used to represent any segment along the backbone of $\alpha^{th}$ chain of kind $j$ 
so that $s_{j,\alpha} \in [0,N_j]$.
Furthermore, each segment on a chain is assigned an electric dipole of length $p_j$ so that the dipole 
moment is of magnitude $e p_j$ ($e$ being the electronic charge), 
where $j = A,B$. Although we have not taken into account the 
variations in the magnitude of these electric dipole moments, but we allow the dipoles to be aligned depending on the 
local electric field. The physical origin of these electric dipoles is the polar nature of the monomers used to synthesize the polymers such as poly(ethylene oxide) and bipolymers like polypeptoids\cite{peptoid_rev}. 
For example, a difference in electronegativities of oxygen and carbon leads to a 
dipole moment of $1.9$ Debye for ethylene oxide monomer\cite{intermolecular_forces}. An important point regarding the origin of electric dipoles is worth mentioning here. 
In addition to the inherent dipole moments of the molecules, finite polarizability of molecules leads to induced 
electric moments\cite{onsager_moments, dielectric} in the presence of other molecules. This induction effect, in turn, enhances\cite{onsager_moments, dielectric} the net 
electric dipole moments. 
In this work, as a first step we study the effects of ``permanent dipole moments'' only and neglect the 
effects of ``induced dipole moments'' which is more relevant 
for highly polarizable or polymers with no permanent moments. 

The thermodynamic properties of a two-component dipolar polymeric system modeled as above are derived from 
the partition function defined through an effective coarse-grained Hamiltonian\cite{edwards_1965,edwardsbook} and 
by following the now-standard field theoretic technique\cite{leibler,ohta_kawasaki,monica_edwards,MM_1993,
zhengwang_blends,fredbook}. The partition function $Z$ is written as a functional path integral over various allowed conformations of all chains and orientations of all segments as

\begin{eqnarray}
       Z & = & \frac {1}{n_A!n_B!}\int \prod_{j=A,B}\prod_{\alpha=1}^{n_j} D[\mathbf{R}_{j,\alpha}] \int \prod_{\alpha=1}^{n_j}\prod_{s_{j,\alpha}=0}^{N_j} d\mathbf{u}_{j,\alpha}(s_{j,\alpha}) \exp \left [-H_0\left\{
\mathbf{R}_{j,\alpha}\right\} \right.\nonumber \\
&& \left.  - H_w\left\{\mathbf{R}_{j,\alpha},\mathbf{R}_{j',\alpha'}\right\} 
- H_{dd}\left\{\mathbf{R}_{j,\alpha},\mathbf{u}_{j,\alpha},\mathbf{R}_{j',\alpha'},\mathbf{u}_{j',\alpha'}\right\} 
\right ] \prod_{\mathbf{r}}\mathbf{\delta}\left[\hat{\rho}_{A}(\mathbf{r}) + \hat{\rho}_{B}(\mathbf{r}) - \rho_0\right]\label{eq:parti_melts}.
\end{eqnarray}
Here the Hamiltonian is written by taking into account
the contributions from the chain connectivity (given by $H_0$ in Eq. (~\ref{eq:connectivity}) below), 
the short ranged repulsive hard-core and attractive dispersive interactions other than the permanent dipole-dipole (represented by $H_w$ in Eq. (~\ref{eq:dispersion})) and the long range electrostatic
interactions between the dipolar species (written as $H_{dd}$, which corresponds to the segment-segment interactions). 
Explicitly, the contribution from the 
chain connectivity is modeled as the Wiener measure\cite{edwardsbook,fredbook,freed_1971}

\begin{eqnarray}
      H_0\left\{\mathbf{R}_{j,\alpha}\right\} &=& \sum_{j=A,B}\frac {3}{2 b_j^2}\sum_{\alpha=1}^{n_j} \int_{0}^{N_j} ds_{j,\alpha} \left(\frac{\partial \mathbf{R}_{j,\alpha}(s_{j,\alpha})}{\partial s_{j,\alpha}} \right )^{2} \label{eq:connectivity}.
\end{eqnarray}
$H_w\left\{\mathbf{R}_{j,\alpha},\mathbf{R}_{j,\alpha'}\right\}$
takes into account the energetic contributions arising from short-range dispersion interactions between segments 
of chains indexed as $\alpha$ and $\alpha'$ located at $\mathbf{R}_{j,\alpha}$ and $\mathbf{R}_{j',\alpha'}$, 
respectively. Following the Edwards formulation for a flexible chain\cite{edwardsbook}, we model these interactions by delta functional/point 
interactions as\cite{edwards_1965,ohta_kawasaki} 
\begin{eqnarray}
H_w\left\{\mathbf{R}_{j,\alpha},\mathbf{R}_{j,\alpha'}\right\} &=& \frac{1}{2}\int d\mathbf{r} 
\left[w_{AA}\hat{\rho}_{A}^2(\mathbf{r}) + w_{BB}\hat{\rho}_{B}^2(\mathbf{r})  + 2 w_{AB} \hat{\rho}_{A}(\mathbf{r})
\hat{\rho}_{B}(\mathbf{r})\right] \label{eq:dispersion}
\end{eqnarray}
Here, $w_{AA}, w_{BB}$ and $w_{AB}$ are the well-known excluded volume parameters describing the 
strengths of interactions between $A-A, B-B$ and $A-B$ monomeric pairs, respectively. These parameters 
characterize strength of short-ranged interactions, whose range is 
parameterized by using delta function interaction potentials.
Also, 
$\hat{\rho}_{j}(\mathbf{r})$ represents microscopic number 
density of the monomers of type $j$ at a certain location $\mathbf{r}$ defined as
        \begin{eqnarray}
\hat{\rho}_{j}(\mathbf{r})  &=& \sum_{\alpha=1}^{n_j} \int_{0}^{N_j} ds_{j,\alpha} \, \delta \left[\mathbf{r}-\mathbf{R}_{j,\alpha}(s_{j,\alpha})\right] \quad \mbox{for} \quad j= A,B
\end{eqnarray}

Electrostatic contributions to the Hamiltonian arising from the segment-segment (or dipole-dipole) 
interactions can be written as\cite{kumar_kilbey} 
\begin{eqnarray}
H_{dd} &=& \frac{l_{Bo}}{2}\int d\mathbf{r}\int d\mathbf{r}' \int d\mathbf{u}\int d\mathbf{u}'
\hat{P}(\mathbf{r},\mathbf{u} )V_{dd}\left(\mathbf{r},\mathbf{u},\mathbf{r}',\mathbf{u}'\right)\hat{P}(\mathbf{r}',\mathbf{u}') \label{eq:segelecideal}  
\end{eqnarray}
where $l_{Bo} = e^2/\epsilon_o k_B T$ is the Bjerrum length in vacuum,  $\epsilon_o$ is the
permittivity of vacuum and $k_BT$ is the Boltzmann constant times the absolute temperature. The dipolar potential 
$V_{dd}$ is given by\cite{intermolecular_forces,MM_dipoles,kumar_glenn}
\begin{eqnarray}
V_{dd}\left(\mathbf{r},\mathbf{u},\mathbf{r}',\mathbf{u}'\right) &=& \frac{\mathbf{u}.\mathbf{u}'}{|\mathbf{r} - \mathbf{r}'|^3}
- 3\frac{\left\{\mathbf{u}.\left(\mathbf{r} - \mathbf{r}'\right)\right\}\left\{\mathbf{u}'.\left(\mathbf{r} - \mathbf{r}'\right)\right\}}{|\mathbf{r}
- \mathbf{r}'|^5} \label{eq:dipolar_potential}
\end{eqnarray}
and $\hat{P}(\mathbf{r},\mathbf{u} ) = p_A \bar{\rho}_A(\mathbf{r},\mathbf{u} ) + p_B \bar{\rho}_B(\mathbf{r},\mathbf{u} )$ is the local dipole density
at $\mathbf{r}$ in the direction of the unit vectror specified by unit vector  $\mathbf{u}$ centered at the location of $\mathbf{r}$.
Here, we have defined a microscopic number density of dipoles of type $A$ and $B$ by $\bar{\rho}_A(\mathbf{r},\mathbf{u})$ 
and $\bar{\rho}_B(\mathbf{r},\mathbf{u})$, respectively. Physically, these functions describe the number of dipoles with 
their centers at a certain location $\mathbf{r}$ with their axes oriented along the unit vector $\mathbf{u}$. 
Formally, these are defined as
\begin{eqnarray}
    \bar{\rho}_{j}(\mathbf{r},\mathbf{u})  &=& \sum_{\alpha=1}^{n_j} \int_{0}^{N_j} ds_{j,\alpha} \, \delta \left[\mathbf{r}
-\mathbf{R}_{j,\alpha}(s_{j,\alpha})\right]
\delta \left[\mathbf{u}-\mathbf{u}_{j,\alpha}(s_{j,\alpha})\right] \quad \mbox{for} \quad j=A,B
\end{eqnarray}

As with the Edwards Hamiltonian for short range inter-segment interactions leading to divergences at short length scales (ultraviolet divergence), the dipolar potential (Eq. ~\ref{eq:dipolar_potential}) also diverges at $|\mathbf{r}-\mathbf{r'}|\rightarrow 0$. We have regularized these divergences by introducing a short range cut-off length of the order 
of the Kuhn segment length. 

Furthermore, we have imposed the incompressibility constraint of the system with the use of the delta function in Eq. ~\ref{eq:parti_melts}. The factorials in Eq. ~\ref{eq:parti_melts} represent the indistinguishability of the chains for each component.

\subsection{Freely rotating dipoles : spinodal, structure factor and interfacial tension}
At high temperatures or for small dipole moments (so that $l_{Bo}p_j \ll 1$ for $j=A,B$), dipoles can rotate 
freely\cite{intermolecular_forces}. 
The condition of free rotation by segmental dipoles is representative of a disordered melt at high temperatures. 
For this particular case, the structure factor and the spinodal curves corresponding to the limits of thermodynamic stability 
can be computed in a relatively straightforward manner as described below. For freely rotating 
dipoles ($l_{Bo}p_j \ll 1$), we can use a series expansion in powers of dipole moments and integrate out the 
orientational degrees of freedom. Truncating the series after fourth power of the dipole moments 
and exponentiating the series\cite{intermolecular_forces,kumar_glenn}, 
the partition function for freely rotating dipoles $Z = \hat{Z}$ (cf. Eq. ~\ref{eq:parti_melts}) can be written as   
\begin{eqnarray}
       \hat{Z} & = & \frac {(4\pi)^{\sum_{j=A,B}n_j N_j}}{n_A!n_B!}\int \prod_{j=A,B}\prod_{\alpha=1}^{n_j} D[\mathbf{R}_{j,\alpha}] 
\exp \left [-H_0\left\{
\mathbf{R}_{j,\alpha}\right\} \right.\nonumber \\
&& \left.  - H_w\left\{\mathbf{R}_{j,\alpha},\mathbf{R}_{j,\alpha'}\right\} 
- \hat{H}_{dd}\left\{\mathbf{R}_{j,\alpha},\mathbf{R}_{j',\alpha'}\right\} 
\right ] \prod_{\mathbf{r}}\mathbf{\delta}\left[\hat{\rho}_{A}(\mathbf{r}) + \hat{\rho}_{B}(\mathbf{r}) - \rho_0\right]
\label{eq:parti_rotating}
\end{eqnarray}
where the angularly averaged dipolar interaction term, $\hat{H}_{dd}$, is given by
\begin{eqnarray}
\hat{H}_{dd} &=& -\frac{l_{Bo}^2}{12}\int d\mathbf{r}\int d\mathbf{r}'
\frac{\left[p_A^2 \hat{\rho}_A(\mathbf{r}) + p_B^2 \hat{\rho}_B(\mathbf{r})\right]\left[p_A^2 \hat{\rho}_A(\mathbf{r}') + p_B^2 \hat{\rho}_B(\mathbf{r}')\right]}{|\mathbf{r} - \mathbf{r}'|^6} 
\label{eq:segelecidealrotating}  
\end{eqnarray}

The dipole interaction in the limit of free rotation becomes essentially short ranged\cite{intermolecular_forces,MM_2004}. 
Within the random phase approximation\cite{leibler,ohta_kawasaki,monica_edwards,zhengwang_blends,sanchez,tang_freed}, the free 
energy up to the fourth degree terms in the order parameter $C(\mathbf{r}) = \hat{\rho}_A(\mathbf{r})/\rho_0 - \phi_A
 = 1-\phi_A - \hat{\rho}_B(\mathbf{r})/\rho_0$ ($\phi_A = n_A N_A/\rho_o V$ being the average volume 
fraction of $A$ monomers, where $\rho_0$ being the average density and $V$ is the total volume  of the homogeneous system ) is given by 
\begin{eqnarray}
\exp\left[-\frac{(\hat{F} - \hat{F}_H)}{k_B T}\right] &=& 
\int \prod_{k\neq 0 }D\left[C_k\right]\exp\left[-\frac{\delta F\{C_k\}}{k_B T}\right] \label{eq:rpa_parti}
\end{eqnarray}
where
\begin{eqnarray}
\frac{\delta F\{C_k\}}{k_B T} &=& \frac{\rho_0}{2}\sum_{k\neq 0} S^{-1}(k) C_k C_{-k}
 + \frac{\rho_0}{6}\sum_{k_1,k_2} \Gamma_3(k_1,k_2,-k_1-k_2) C_{k_1}C_{k_2}C_{-k_1-k_2} \nonumber \\
&& + \frac{\rho_0}{12}\sum_{k_1,k_2,k_3} \Gamma_4(k_1,k_2,k_3,-k_1-k_2-k_3) 
C_{k_1}C_{k_2}C_{k_3}C_{-k_1-k_2-k_3}
\label{eq:vertex_funcs}
\end{eqnarray}
In these expressions, $k,k_j$ are the wavevectors and $C_k$ is the Fourier component of $C(\mathbf{r})$. $\hat{F}_H$ is the free energy of the homogeneous phase (corresponding to $k=0$), 
given by
\begin{eqnarray}
\frac{\hat{F}_H }{\rho_o V k_B T} &=& \frac{\rho_0}{2}\left[w_{AA}\phi_A + w_{BB}(1-\phi_A)\right] + \frac{\phi_A }{N_A}\left[\ln \left(\frac{\phi_A}{N_A}\right) -1\right] + \frac{(1-\phi_A) }{N_B}\left[\ln \left(\frac{1-\phi_A}{N_B}\right)-1\right] \nonumber \\
&& +  \chi_{AB} \phi_A (1-\phi_A) 
- \frac{\pi \rho_o l_{Bo}^2}{9\lambda^3}\left[p_A^2\phi_A + p_B^2 (1-\phi_A)\right]^2 -\ln 4\pi \label{eq:free_dipoles}
\end{eqnarray} 
where  $\lambda$ is a cut-off length below which $1/r^6$ dipolar 
potential is no longer valid. Also, we have defined 
\begin{eqnarray}
\chi_{AB}b^3 &=& w_{AB} - \frac{w_{AA} + w_{BB}}{2} \label{eq:chi_parameter_main}
\end{eqnarray}
and used $\rho_0 b^3 = 1$ in writing Eq. ~\ref{eq:free_dipoles}. 
Spinodal curve can be computed using Eq. ~\ref{eq:free_dipoles} and is given by 
\begin{eqnarray}
\frac{\partial^2 }{\partial \phi_A^2}\left[\frac{\hat{F}_H }{\rho_o V k_B T}\right] &=& \frac{1}{\phi_A N_A} + \frac{1}{(1-\phi_A) N_B}
-2\chi_{AB} - \frac{2\pi \rho_o l_{Bo}^2}{9\lambda^3}\left[p_A^2 - p_B^2 \right]^2 = 0 \label{eq:spino_free_dipoles}
\end{eqnarray}
It is clear from this expression that dipolar interactions renormalizes ``chi'' parameter 
and we can write an ``effective'' parameter as 
\begin{eqnarray}
\chi_{AB,eff} &=& \chi_{AB} + \frac{p^4}{16\pi \rho_0\lambda^3} \quad \mbox{where} \quad p^2 = \frac{4\pi}{3} l_{Bo} \rho_0 \left(p_A^2-p_B^2\right) \label{eq:chi_eff}
\end{eqnarray}
, which is in accord with the well-known law,``polar likes polar''. Also, note that the Edwards' approach
for using a delta function potential in order to model short range interactions along with the local
incompressibility constraint naturally leads to appearance of the dimensionless Flory's chi parameter $\chi_{AB}$ in the theory, defined in
terms of excluded volume parameters $w_{AA},w_{BB}$ and $w_{AB}$ by Eq. ~\ref{eq:chi_parameter_main}. 
The physical origin of the excluded volume parameters\cite{scalingbook}
is the repulsive hard-core and attractive dispersive interactions and in general, $w_{ij} =w_{ij}^0 -\omega\alpha_i \alpha_j$ so that $\omega>0$ (the 
dispersive interactions being
attractive), $\alpha_i$ is the electronic polarizability of molecule $i$ and $w_{ij}^0$ is the volume excluded 
due to hard-cores. Using the definition of $\chi_{AB}$ in terms of excluded volume parameters, it has been postulated\cite{scalingbook} that 
$\chi_{AB}b^3 = \chi_{AB}^0 b^3 + \omega (\alpha_A - \alpha_B)^2/2 >0$ where 
$\chi_{AB}^0 b^3 = w_{AB}^0 - \frac{w_{AA}^0 + w_{BB}^0}{2}$.
Additional contribution in $\chi_{AB,eff}b^3$ arising from dipolar
interactions is in qualitative agreement with such a description of the ``chi'' parameter in terms of
electronic polarizabilities if we invoke the relation for total polarizability of a molecule of type $j$ as
$\alpha_j = \alpha_o + p_j^2/3k_{B}T$ (known as the Debye-Langevin\cite{intermolecular_forces} equation, which is strictly valid at high temperatures), where $\alpha_o$ is the electronic
polarizability and the rest is the orientational polarizability.

For uncharged polymers, the spinodal `chi' parameter $\chi_{s0}$ follows from Eq. ~\ref{eq:spino_free_dipoles} as
\begin{equation}
2 \chi_{s0}=\frac {1}{\phi_A N_A} + \frac {1}{(1-\phi_A)N_B}. 
\end{equation}
Writing this as being proportional to the inverse of temperature $T_{s0}$, and noting that the 
Bjerrum length 
$l_{B0}$ is inversely proportional to $T$, the spinodal temperature $T_s$ for the presence of dipole moment mismatch 
follows from Eq. ~\ref{eq:spino_free_dipoles} as 
\begin{equation}
\frac {T_s}{T_s0} = \frac {1}{2} \left [ 1+ \left\{1+\frac {4 \pi \rho_0}{9 \chi_{s0} \lambda^3} \left(\frac {e^2}{\epsilon_0 k_BT_{s0}}\right)^2 (p_A^2-p_B^2)^2\right\}^{1/2}\right].
\end{equation}
Therefore, the spinodal temperature is increased by the mismatch in the dipole moments of the components. This result based on freely rotating dipolar segments turns out to be a general result as shown below.

Furthermore, effects of dipolar interactions on the structure factor and correlation length can be studied by 
analyzing the second order term in Eq. ~\ref{eq:vertex_funcs}, which is the inverse of the structure factor, $S^{-1}(k)$, 
given by 
\begin{eqnarray}
S^{-1}(k)& = & S_{bare}^{-1}(k)+  S_{dd}^{-1}(k) \label{eq:struct}\\
\mbox{so that}\quad
S_{bare}^{-1}(k)&=& \frac{1}{\phi_A N_A g(x_A)} + \frac{1}{(1-\phi_A) N_B g(x_B)}- 2\chi_{AB} \\
\mbox{and}\quad S_{dd}^{-1}(k)&= & -\frac{p^4}{8 \pi \rho_0} \left[\frac{1}{\lambda^3} - \frac{k^2}{2\lambda} + \frac{\pi k^3}{16}\right] \label{eq:ddstruct}
\end {eqnarray}
In Eq. ~\ref{eq:struct}, $S_{bare}^{-1}(k)$ is the contribution due to Gaussian chains interacting 
via short range excluded volume interactions
and $S_{dd}^{-1}(k)$ is the additional effect due to the dipolar interactions present in
the system. In writing Eq. ~\ref{eq:ddstruct}, we have used mathematical expression for 
Fourier Transform (FT) of $1/r^6$, given by
$\mbox{FT}(1/r^6) = 4\pi \int_0^{\infty}dr \frac{\sin kr}{kr^{5}} = \frac{4\pi}{3}\left[\frac{1}{\lambda^3} - \frac{k^2}{2\lambda} + \frac{\pi k^3}{16}\right]$ so that 
$\lambda \rightarrow 0$ is the cutoff length included to regularize divergent integrals.
Also, $g(x_j)$ is the Debye function\cite{edwardsbook,fredbook} given by
\begin{equation}
g(x_j)= \frac{2(e^{-x_j}-1 + x_j)}{x_j^{2}},\quad x_j = \frac{k^{2}N_j b_j^{2}}{6}= k^{2}R_{g,j}^{2}  
\label{eq:debye}
\end{equation}
The correlation length\cite{scalingbook} $\xi$ in dipolar polymer blends can be extracted from Eq. ~\ref{eq:struct} by using an expansion\cite{scalingbook,edwardsbook} of the Debye function at small wavevectors $k$ 
($g(x_j)\rightarrow 1 - x_j/3$ for $x_j<1$). 
The approximation for the Debye function lets us write 
\begin{eqnarray}
S^{-1}(k)& = & \frac{1}{\phi_A N_A } + \frac{1}{(1-\phi_A) N_B }- 2\chi_{AB,eff}  
   + \frac{k^2}{18}\left[\frac{b_A^2}{\phi_A} + \frac{b_B^2}{1-\phi_A} + \frac{9 p^4}{8\pi \rho_0 \lambda} \right]  
\label{eq:invsk_near_bulk}
\end {eqnarray}
where we have kept terms up to $k^2$. Writing $S(k) = S(0)/(1 + k^2 \xi^2)$, $\xi$ is given by
\begin{eqnarray}
\xi & = & \frac{1}{\sqrt{18}}\left[ \frac{b_A^2}{\phi_A} + \frac{b_B^2}{1-\phi_A} + \frac{9p^4}{8\pi \rho_0 \lambda} 
\right]^{1/2}\left[\left|\frac{1}{\phi_A N_A } + \frac{1}{(1-\phi_A) N_B } -2\chi_{AB,eff}\right|\right]^{-1/2}
\label{eq:xi}
\end {eqnarray}
From Eqs. ~\ref{eq:chi_eff} and ~\ref{eq:xi}, it is clear that the correlation length increases with an 
increase in mismatch between the monomeric dipole moments. Also, using Eqs. ~\ref{eq:spino_free_dipoles} and 
~\ref{eq:xi}, the correlation length $\xi $
diverges at the critical point $T = T_c$ with the exponent of $-1/2$.

Effects of dipolar interactions on the correlation length manifest in the density profiles as well as 
interfacial tension of the interface
between coexisting phases near the critical point. 
Using Eqs. ~\ref{eq:rpa_parti} and ~\ref{eq:vertex_funcs}, we have 
studied interfacial tension between co-existing phase near the critical point within the saddle-point approximation. 
Approximating $S^{-1}(k)$ by Eq. ~\ref{eq:invsk_near_bulk}, $\Gamma_3(k_1,k_2,k_3)$ and $\Gamma_4(k_1,k_2,k_3,k_4)$ by their value\cite{sanchez,tang_freed} 
at $|k_j|=0$, we have 
constructed density profiles and the interfacial tension. Using such a procedure leads to 
a volume fraction profile\cite{cahn_hilliard,joanny_leibler} given by 
\begin{eqnarray}
\phi_A(z) &=& \phi_{A,c} + \frac{\left[\phi_{A}(\infty) - \phi_{A}(-\infty)\right]}{2} \tanh \left[\frac{z}{\sqrt{2}\xi_c}\right] \label{eq:cahn_density}
\end{eqnarray}
where $z$ is the direction perpendicular to the interface and $\phi_{A,c}$ is the volume fraction of $A$ at the critical point. Furthermore,  $\phi_{A}(\pm\infty) = \phi_{A,c} \pm \sqrt{3|\Gamma_2(0)|/\Gamma_4(0)}$ so that 
$\Gamma_2(0) = \left[S^{-1}(0)\right]_{\phi_A = \phi_{A,c}}, \xi_c = \left[\xi\right]_{\phi_A = \phi_{A,c}}$
and   
\begin{eqnarray}
\Gamma_4(0) & = & \frac{1}{\phi_{A,c}^3 N_A } + \frac{1}{(1-\phi_{A,c})^3 N_B }
\end {eqnarray}
It is to be noted that the volume fraction in the phase at $z \rightarrow \infty$ tends to increase with 
an increase in the mismatch between the monomeric dipole moments. In other words, an increase in the mismatch tends to 
promote phase segregation. From Eq. ~\ref{eq:free_dipoles}, the volume fraction at the critical point is given by  
\begin{eqnarray}
\phi_{A,c} &=& \frac{\sqrt{N_B}}{\sqrt{N_A} + \sqrt{N_B}}
\end {eqnarray}
and stays the same as in the absence of dipolar interactions. In the next section, we will show that this is a direct 
consequence of the neglect of orientational effects in the case of freely rotating dipoles and in fact, critical point gets shifted by dipolar interactions.  

Nevertheless, using Eq. ~\ref{eq:cahn_density} for the volume fraction profile, interfacial tension near the critical point, $\sigma_c$, (per unit area) can be 
readily computed, given by 
\begin{eqnarray}
\frac{\sigma_c}{A k_B T} &=& \frac{2\sqrt{2}\Gamma_2(0)^2 \xi_c}{\Gamma_4(0)}
= \frac{2}{3}\frac{\left|\Gamma_2(0)\right|^{3/2}}{\Gamma_4(0)} \left[ \frac{b_A^2}{\phi_{A,c}} 
+ \frac{b_B^2}{1-\phi_{A,c}} + \frac{9p^4}{8\pi \rho_0 \lambda} \right]^{1/2} \label{eq:cahn_interfacial}
\end {eqnarray}
where $A$ is the area of the planar interface. 

From Eq. ~\ref{eq:cahn_interfacial}, it is clear that the interfacial tension increases with an increase in 
mismatch between the dipole moments of the two monomers in polymer blends. 
Also, Eqs. ~\ref{eq:xi} and ~\ref{eq:cahn_density} reveal that interfacial width ($=\sqrt{2}\xi_c$) 
\textit{increases} with an 
increase in the mismatch i.e., interfaces become more diffuse with an increase in the mismatch. 
Note that the approximation of freely rotating dipoles is stictly valid 
in the high temperature limit close to one phase regime and it is clear that such an 
approximation will be invalid in the two phase regime. In order to study inhomogeneous structures such as a 
sharp interface between coexisting phases, we need 
to construct a theory capable of taking into account the effects of inhomogeneities on the dipolar interactions. In the 
following, we present such a general formalism using field theoretical transformations\cite{fredbook} 
and use it to study coexistence curves as well as the interfacial tension. Predictions of the field theory are contrasted with 
the predictions presented above for freely rotating dipoles.  
 
\subsection{Field theory}

Before introducing fields, we can write the electrostatic terms in a more convenient form by defining a quantity
$\hat{P}_{ave}(\mathbf{r}) = \int d\mathbf{u} \hat{P}(\mathbf{r},\mathbf{u})\mathbf{u}$, which is a vector.
$H_{dd}$ in Eq. ~\ref{eq:segelecideal} can be written in terms of $\hat{P}_{ave}(\mathbf{r})$ as
\begin{eqnarray}
H_{dd} &=& \frac{l_{Bo}}{2}\int d\mathbf{r}\int d\mathbf{r}' 
\frac{\left[\nabla_{\mathbf{r}}.\hat{P}_{ave}(\mathbf{r}) \right]\left[\nabla_{\mathbf{r}'}.\hat{P}_{ave}(\mathbf{r}')
\right]}{|\mathbf{r} - \mathbf{r}'|} \label{eq:final_particle_elec}
\end{eqnarray}

Using field theoretical transformations described in Appendix A, we can write the partition function for the blends 
as 
\begin{eqnarray}
       Z & = & \frac{Z_o}{\mu_\psi}\int D\left[\rho_A\right]\int  D\left[w_A\right]\int  D\left[w_B\right]\int  D\left[\psi\right]
\exp \left [- \frac{H}{k_BT}\right ] \label{eq:parti_physical}
\end{eqnarray}
where $Z_o = \exp(-\frac{F_0}{k_BT})$ is the partition function for the same system in the absence of the 
interactions, given by
\begin{eqnarray}
      \frac{F_0}{k_BT} &=& \frac{1}{2}\left[w_{AA}n_A N_A + w_{BB}n_B N_B\right]\rho_0 
+ \ln \left[\frac{n_A! n_B!}{(4\pi)^{n_A N_A + n_B N_B} V^{n_A + n_B}}\right] 
\end{eqnarray}
and
\begin{eqnarray}
       \frac{H}{k_BT} &=& \chi_{AB}b^3\int d\mathbf{r} \rho_A(\mathbf{r})(\rho_0 - \rho_A(\mathbf{r})) - i \int d\mathbf{r} w_A(\mathbf{r})\rho_A(\mathbf{r}) 
- i \int d\mathbf{r} w_B (\mathbf{r})\left[\rho_0 - \rho_A(\mathbf{r}) \right] 
\nonumber \\
&& - \frac{1}{8\pi l_{Bo}}\int d\mathbf{r} \psi(\mathbf{r})\nabla_{\mathbf{r}}^2 \psi(\mathbf{r}) - \int d\mathbf{r}\rho_A(\mathbf{r})\ln \left[\frac{\sin\left(p_A |\nabla_\mathbf{r}\psi(\mathbf{r})|\right)}
{p_A |\nabla_\mathbf{r}\psi(\mathbf{r})|}\right] 
\nonumber \\
&& - \int d\mathbf{r} \left\{\rho_0 - \rho_A(\mathbf{r})\right\} 
\ln \left[\frac{\sin\left(p_B |\nabla_\mathbf{r}\psi(\mathbf{r})|\right)}
{p_B |\nabla_\mathbf{r}\psi(\mathbf{r})|}\right]  - \sum_{j=A,B}n_j\ln \bar{Q}_{j}\left\{w_j\right\}\label{eq:hami_physical}
\end{eqnarray}
Here, $\bar{Q}_{j}$ is \textit{normalized} partition functions for a polymer chain of type $j$ described explicitly 
in Appendix A. In these equations, $\psi$ is the collective field introduced to decouple electrostatic interactions and is the equivalent of electrostatic 
potential. Also, $w_{A}, w_B$ are the fields 
introduced to decouple short range interactions modeled by Edwards' delta functional approach\cite{fredbook,edwardsbook}. Furthermore, $\rho_A$ is collective variable characterizing local density of monomers of type $A$.
 
It is clear from the partition function (cf. Eqs. ~\ref{eq:parti_physical} and ~\ref{eq:hami_physical}) that the
parameters $p_A|\nabla_\mathbf{r}\psi(\mathbf{r})|$
and $p_B|\nabla_\mathbf{r}\psi(\mathbf{r})|$ are the
parameters controlling the magnitude of the effect of electric dipoles on the thermodynamics.
An important insight into the effect of electric dipoles on the thermodynamics
of the blends is obtained if one considers the weak coupling limit\cite{kumar_kilbey} for the electric dipoles (i.e.,
the limit of weak dipoles or weak local electric fields) so that
$p_A |\nabla_\mathbf{r}\psi(\mathbf{r})|\rightarrow 0$, $p_B|\nabla_\mathbf{r}\psi(\mathbf{r})| \rightarrow 0$ 
(note that the ``weak coupling limit'' considered in this work is different from 
the one used in literature\cite{netz_coupling,naji_coupling} on electrolytes characterized by the coupling constant 
$z\psi$, $z$ being the charge). In this limit, we can use the approximation $\ln \left[\sin x /x\right] \simeq -x^2/6$ in the expression 
for the partition function.
Using the approximations, it can be shown that the electric dipoles renormalize the Bjerrum length of the medium by the relation
\begin{eqnarray}
\frac{1}{l_B(\mathbf{r})} &=& \frac{1}{l_{Bo}} + \frac{4\pi}{3}\left[p_A^2 \rho_A(\mathbf{r}) + p_B^2 \left(\rho_0 - \rho_A(\mathbf{r})\right)\right] \label{eq:bjerrum_weak}
\end{eqnarray}
where $l_B(\mathbf{r})$ is the renormalized Bjerrum length in the polymer blend. One can also use this relation
to define an effective local dielectric function if we use the relation between the Bjerrum length and the dielectric constant
as $l_B(\mathbf{r}) = e^2/\epsilon_o \epsilon(\mathbf{r})k_B T$. This allows us to write,
\begin{eqnarray}
\epsilon(\mathbf{r}) &=& \epsilon_{A}\phi_A(\mathbf{r}) + \epsilon_B \phi_B(\mathbf{r}),\label{eq:linearmix}
\end{eqnarray}
where $\phi_A(\mathbf{r}) = \rho_A(\mathbf{r})/\rho_0$ and $\phi_B(\mathbf{r}) = (\rho_0 - \rho_A(\mathbf{r}))/\rho_0$ 
are
the volume fractions of A and B monomers, respectively. Furthermore,
 \textit{relative} dielectric ``constant'' of the pure components (with respect to vacuum) 
are given by the relations
\begin{eqnarray}
\epsilon_A &=& 1 + \frac{4\pi l_{Bo}}{3} p_A^2\rho_0 \label{eq:dielec_pure}\\
\epsilon_B &=& 1 + \frac{4\pi l_{Bo}}{3} p_B^2\rho_0. \label{eq:dielec_pure2}
\end{eqnarray}
Note that Eq. ~\ref{eq:linearmix} for the local dielectric function is valid, in general, for the 
homogeneous  as well as 
inhomogeneous systems. In the subsequent section, we present the implications of the concentration 
dependence and inhomogeneous nature of the dielectric function 
on the thermodynamics of polymer blends. In the weak coupling limit, functional integral 
over $\psi$ can be computed exactly for one-dimensional
variation in densities e.g., in the case of a planar interface between the co-existing phases. 
In the following, we present details of method of evaluating the
functional integral. After the evaluation, we approximate other functional
integrals by the saddle-point approximation to construct a free energy expression.
Due to the fact that the functional integral over $\psi$ is computed without invoking 
the saddle-point approximation, the results presented in this work goes beyond the 
mean-field description of electrostatics. 

\subsection{Electrostatic contribution}
We start by rewriting Eq. ~\ref{eq:parti_physical} as 
\begin{eqnarray}
       Z & = & Z_o \int D\left[\rho_A\right]\int  D\left[w_A\right]\int  D\left[w_B\right]
\exp \left [- \frac{H_{neu}-H_{elec}}{k_BT}\right ] \label{eq:parti_physical2}
\end{eqnarray}
where 
\begin{eqnarray}
       \frac{H_{neu}}{k_BT} &=& \chi_{AB}b^3\int d\mathbf{r} \rho_A(\mathbf{r})(\rho_0 - \rho_A(\mathbf{r})) - i \int d\mathbf{r} w_A(\mathbf{r})\rho_A(\mathbf{r}) 
- i \int d\mathbf{r} w_B (\mathbf{r})\left[\rho_0 - \rho_A(\mathbf{r}) \right] 
\nonumber \\
&&- \sum_{j=A,B}n_j\ln \bar{Q}_{j}\left\{w_j\right\}\label{eq:hami_physical2}
\end{eqnarray}
and
\begin{eqnarray}
\frac{H_{elec}}{k_B T} &=& -\ln\left[\frac{\int D[\psi] \exp\left(-\frac{1}{8\pi l_{Bo}}\int d\mathbf{r} \epsilon(\mathbf{r})|\nabla_{\mathbf{r}}\psi(\mathbf{r})|^2\right)}{\int D[\psi] \exp\left(-\frac{1}{8\pi l_{Bo}}\int d\mathbf{r} |\nabla_{\mathbf{r}}\psi(\mathbf{r})|^2\right)}\right]
\end{eqnarray}
and $\epsilon(\mathbf{r})$ is given by Eq. ~\ref{eq:linearmix}. 
In the following, we rewrite Eq. ~\ref{eq:linearmix} 
as  
\begin{eqnarray}
\epsilon(\mathbf{r}) &=& \epsilon_B + p^2 \rho_A(\mathbf{r})/\rho_o\label{eq:bjerrum_wcl}\\
\mbox{so that} \quad p^2 &=& \frac{4\pi l_{Bo}\rho_o}{3}(p_A^2 - p_B^2) = \epsilon_A - \epsilon_B
\end{eqnarray}
where $\epsilon_A$ and $\epsilon_{B}$ are the dielectric constants for the pure ``A'' and ``B'' phases, respectively. 
(cf. Eqs. ~\ref{eq:dielec_pure} and ~\ref{eq:dielec_pure2}). Also, the parameter $p$ is the same as 
defined in Eq. ~\ref{eq:chi_eff} while considering freely-rotating dipoles. In general, computation of $H_{elec}$ is 
a hard problem due to need to 
compute all eigenvalues of the operator in the exponent. However, for 
one dimensional variation in the volume fraction (and in turn, the dielectric function), 
$H_{elec}/k_B T$ can be computed exactly by writing it as 
\begin{eqnarray}
\frac{H_{elec}}{k_B T} &=& -\ln\left[\frac{\int D[\psi] \exp\left(\frac{1}{8\pi l_{Bo}}\int d\mathbf{r} 
\psi(\mathbf{r})\nabla_{\mathbf{r}}.\left\{\epsilon(\mathbf{r})\nabla_{\mathbf{r}}\psi(\mathbf{r})\right\}\right)}{\int D[\psi] \exp\left(\frac{1}{8\pi l_{Bo}}\int d\mathbf{r} \psi(\mathbf{r})\nabla_{\mathbf{r}}^2\psi(\mathbf{r})\right)}\right]\label{eq:felec_real}
\end{eqnarray}
where we have used integration by parts and the fact that $\nabla_{\mathbf{r}}\psi(\mathbf{r}) = 0 $ at the boundaries. 
For one-dimensional variations in dielectric function so that $\epsilon(\mathbf{r})\equiv \epsilon(z)$, 
we can use two-dimensional Fourier-Transform for $\psi(\mathbf{r})$ defined by 
$\psi(\mathbf{r}) = \sum_{\mathbf{q}} \hat{\psi}_{\mathbf{q}}(z) \exp^{i\mathbf{q}.\mathbf{r_{\parallel}}}
, \mathbf{r_{\parallel}}$ being an in-plane position vector. 
Using the Fourier-Transform, Eq. ~\ref{eq:felec_real} becomes
\begin{eqnarray}
\frac{H_{elec}}{k_B T} &=& -\sum_{\mathbf{q}}\ln\left[\frac{\int D[\hat{\psi}_{\mathbf{q}}(z)] \exp\left(-\frac{1}{2}\int_{0}^{L} dz \hat{\psi}_{\mathbf{q}}(z)M_{\mathbf{q}}(z)\hat{\psi}_{-\mathbf{q}}(z)\right)}
{\int D[\hat{\psi}_{\mathbf{q}}(z)] \exp\left(-\frac{1}{2}\int_{0}^{L} dz \hat{\psi}_{\mathbf{q}}(z)\hat{M}_{\mathbf{q}}(z)\hat{\psi}_{-\mathbf{q}}(z)\right)}\right]\label{eq:felec_fourier}
\end{eqnarray}
where 
\begin{eqnarray}
M_{\mathbf{q}}(z) &=& \frac{\epsilon(z)}{4\pi l_{Bo}}\left[-\frac{\partial^2}{\partial z^2} + q^2 
- \frac{\partial \ln \epsilon(z)}{\partial z}\frac{\partial }{\partial z}\right]\\
\hat{M}_{\mathbf{q}}(z) &=& \frac{1}{4\pi l_{Bo}}\left[-\frac{\partial^2}{\partial z^2} + q^2\right]
\end{eqnarray}
so that $L$ is the length of region under consideration, which will be extended to infinity at appropriate place. 
Now, using identity 
for an arbitrary function $\hat{f}(z)$, 
\begin{eqnarray}
M_{\mathbf{q}}(z)\hat{f}(z) &=& \frac{\sqrt{\epsilon(z)}}{4\pi l_{Bo}}\left[-\frac{\partial^2}{\partial z^2} + q^2 
+ \frac{1}{\sqrt{\epsilon(z)}}\frac{\partial^2 \sqrt{\epsilon(z)}}{\partial z^2}\right]\left(\sqrt{\epsilon(z)} \hat{f}(z)\right)
\end{eqnarray}
we can write Eq. ~\ref{eq:felec_fourier} as  
\begin{eqnarray}
\frac{H_{elec}}{k_B T} &=& \frac{1}{2 \Delta z}\sum_{\mathbf{q}}\int_{0}^L dz \ln \epsilon(z) - \sum_{\mathbf{q}}\ln\left[\frac{\int D[\bar{\psi}_{\mathbf{q}}(z)] \exp\left(-\frac{1}{2}\int_{0}^{L} dz \bar{\psi}_{\mathbf{q}}(z)\bar{M}_{\mathbf{q},1}(z)\bar{\psi}_{-\mathbf{q}}(z)\right)}
{\int D[\bar{\psi}_{\mathbf{q}}(z)] \exp\left(-\frac{1}{2}\int_{0}^{L} dz \bar{\psi}_{\mathbf{q}}(z)\bar{M}_{\mathbf{q},0}(z)\bar{\psi}_{-\mathbf{q}}(z)\right)}\right]\nonumber \\
&&\label{eq:felec_fourier_final}
\end{eqnarray}
where $\Delta z$ is the grid spacing used to discretize the space, 
$\bar{\psi}_{\mathbf{q}}(z) = \sqrt{\epsilon(z)}\hat{\psi}_{\mathbf{q}}(z)$
and 
\begin{eqnarray}
\bar{M}_{\mathbf{q},1}(z) &=& \left[-\frac{\partial^2}{\partial z^2} + \bar{q}^2(z)\right]\\
\bar{M}_{\mathbf{q},0}(z) &=& \left[-\frac{\partial^2}{\partial z^2} + q^2\right]
\end{eqnarray}
so that $\bar{q}^2(z)  = q^2 + \frac{1}{\sqrt{\epsilon(z)}}\frac{\partial^2 \sqrt{\epsilon(z)}}{\partial z^2}$.
Now, evaluation of the second term on the right hand side of Eq. ~\ref{eq:felec_fourier_final} can 
be done using methods for computation of functional determinants\cite{kleinertbook}. In particular, 
we use the notation 
\begin{eqnarray}
\frac{H_{elec}}{k_B T} &=& \frac{1}{2\Delta z}\sum_{\mathbf{q}}\int_{0}^L dz \ln \epsilon(z) +\frac{1}{2}\sum_{\mathbf{q}}\ln\left[\frac{Det
\left[\bar{M}_{\mathbf{q},1}\right]}{Det\left[\bar{M}_{\mathbf{q},0}\right]}\right] \label{eq:felec_det}
\end{eqnarray}
and $Det\left[\bar{M}_{\mathbf{q},1}\right]$ is given by the relation 
\begin{eqnarray}
Det\left[\bar{M}_{\mathbf{q},1}\right] &=& G_{\mathbf{q},1}(z=L)
\end{eqnarray}
where function $G_{\mathbf{q},1}$ satisfies
\begin{eqnarray}
\bar{M}_{\mathbf{q},1}(z)G_{\mathbf{q},1}(z) &=& 0 \label{eq:det_eqn}
\end{eqnarray}
with the boundary conditions $G_{\mathbf{q},1}(0) = 0$ and $\left\{\frac{\partial G_{\mathbf{q},1}(z)}
{\partial z}\right\}_{|z=0} = 1$.
Noting that $\bar{M}_{\mathbf{q},0}$ is a special case of $\bar{M}_{\mathbf{q},1}$ and corresponds to 
spatially invariant dielectric function, the same method can be used to compute $Det\left[\bar{M}_{\mathbf{q},0}\right]$. So far, we have mapped the problem of computing infinite eigenvalues for an operator on to a much simpler boundary value problem. 

A general solution of Eq. ~\ref{eq:det_eqn} needs to be computed numerically. However, in the following, we 
use Wentzel-Kramers-Brillouin (WKB) approximation\cite{bender_orszag} to get insights into the effects of dielectric inhomogenities on the free energy of 
polymer blends. Introducing a length scale $R$ (of the order of radius of gyration of polymers) to 
define dimensionless variables 
$\bar{z} = z/R, \bar{L} = L/R, \tilde{q} = qR$, WKB approximation gives an analytical expression for the 
solution of Eq. ~\ref{eq:det_eqn} in the limit of $R \rightarrow \infty$. 
Using the solution, Eq. ~\ref{eq:felec_det} can be written as 
\begin{eqnarray}
\frac{H_{elec}}{k_B T} &=& \frac{1}{2\Delta z}\sum_{\mathbf{q}}\int_{0}^L dz \ln \epsilon(z) + \frac{1}{2}\sum_{\mathbf{q}}\ln\left[\frac{\sinh\left\{\int_{0}^{\bar{L}} d\bar{z} \sqrt{\hat{q}^2(\bar{z})}\right\}}{\sinh \left\{\tilde{q}\bar{L}\right\}}\frac{\tilde{q}}{\sqrt{\hat{q}(0)\hat{q}(\bar{L})}}\right]
\end{eqnarray}
where $\hat{q}^2(\bar{z}) =  \tilde{q}^2 + \frac{1}{\sqrt{\epsilon(\bar{z})}}\frac{\partial^2 \sqrt{\epsilon(\bar{z})}}{\partial \bar{z}^2}$. For a planar interface between co-existing phase, $z=0$  and $z=L$ corresponds to isotropic phases so that $\hat{q}(0) = \hat{q}(\bar{L}) = \tilde{q}$. In the limit of $L\rightarrow \infty, R \rightarrow \infty, \sum_q =  A\int d^2q/(2\pi)^2$ 
and expanding the second term in powers of gradients of dielectric function, we get 
\begin{eqnarray}
\frac{H_{elec}}{k_B T} &=& \frac{A q_{max}^2}{8 \pi \Delta z}\int_{0}^{L} dz \ln \epsilon(z) + 
\frac{A q_{max}}{8\pi} \int_0^{L}dz\frac{1}{\sqrt{\epsilon(z)}}\frac{\partial^2 \sqrt{\epsilon(z)}}{\partial z^2} \label{eq:interface_ssl}
\end{eqnarray}
where $q_{max}$ is the maximum value of wave-vector, introduced in order to regularize divergent integrals, whose 
origin lies in the divergent dipolar potential for $|\mathbf{r}-\mathbf{r}'|\rightarrow 0$. 

For a homogeneous phase, the second term on the right hand side of 
Eq. ~\ref{eq:interface_ssl} vanishes. In the next section, we show the effects of the 
leading term in Eq. ~\ref{eq:interface_ssl} on the phase coexistence curves. 
However, the second term takes into account non-local
effects relevant for an inhomogeneous medium. Using integration by parts for the non-local term, 
we get
\begin{eqnarray}
\frac{H_{elec}}{k_B T} &=& \frac{A q_{max}^2}{8 \pi \Delta z}\int_{0}^{L} dz \ln \epsilon(z) + 
\frac{A q_{max}}{32\pi} \int_0^{L}dz \left[\frac{\partial \ln \epsilon(z)}{\partial z}\right]^2 
\label{eq:final_nonlocal}
\end{eqnarray} 
Such a functional form for the non-local effects of dielectric function allows us to study 
interfacial tension for a planar interface in the strong segregation limit as described below. 
Also, note that the functional form for the non-local effects of the dielectric function is similar to the one 
derived in Ref. \cite{podgornik} by transfer matrix approach. 
  
\section{Results}\label{sec:results}
\subsection{Phase co-existence}
It is well-known\cite{fredbook} that the homogeneous phase is one of the solutions of the saddle-point equations, which
can be derived by optimizing the Hamiltonian given by Eqs. ~\ref{eq:parti_physical2}, ~\ref{eq:hami_physical2} and 
~\ref{eq:interface_ssl} with respect to $w_A,w_B$ and $\rho_A$. Carrying out the optimizations, we obtain
\begin{eqnarray}
\phi_A(\mathbf{r}) = \frac{\rho_A^\star(\mathbf{r})}{\rho_0}  &=& i \left[\frac{n_A}{\rho_0}\frac{\delta \ln \bar{Q}_{A}}{\delta w_A(\mathbf{r})}\right]_{w_A=w_A^\star} \\
1 - \phi_A(\mathbf{r})  &=& i \left[\frac{n_B}{\rho_0}\frac{\delta \ln \bar{Q}_{B}}{\delta w_B(\mathbf{r})}\right]_{w_B = w_B^\star} 
\end{eqnarray}
\begin{eqnarray}
iw_A^\star(\mathbf{r}) -  iw_B^\star(\mathbf{r}) 
&=& \chi_{AB} \left[1 - 2\phi_A(\mathbf{r})\right] + \frac{q_{max}^2 \Delta \epsilon }{8\pi \rho_o \Delta z (1 + \Delta \epsilon 
\phi_A(z))} \nonumber \\
&&- \frac{q_{max}\Delta \epsilon}{16 \pi \rho_0\left[1 + \Delta \epsilon \phi_A(z)\right]}
\frac{\partial^2 \ln \left[1 + \Delta \epsilon \phi_A(z)\right]}{\partial z^2}
\end{eqnarray}
where we have defined $\Delta \epsilon =  (\epsilon_A - \epsilon_B)/\epsilon_B$. 
It can be checked explicitly that
$\phi_A(\mathbf{r}) = \phi_A , w_A^\star = \mbox{constant}$ and $w_B^\star = \mbox{constant}$ satisfy
these saddle-point equations. Furthermore, these parameters represent a homogeneous phase.
The free energy of the homogeneous phase can be obtained by plugging this trivial
solution of the saddle-point equations into the Eq. ~\ref{eq:parti_physical}. 
Explicitly, we obtain the free energy of the
homogeneous phase as
\begin{eqnarray}
      \frac{F_H}{\rho_0 V k_BT} &=& \frac{\rho_0}{2}\left[w_{AA} \phi_A + w_{BB} (1-\phi_A)\right] + \chi_{AB}\phi_A (1-\phi_A) 
+ \frac{\phi_A}{N_A}\left[\ln \left(\frac{\phi_A}{N_A}\right) -1\right] \nonumber \\
&& + \frac{1-\phi_A}{N_B}\left[\ln \left(\frac{1-\phi_A}{N_B}\right) -1\right] - \ln 4\pi 
+  \frac{q_{max}^2}{8\pi \Delta z \rho_o} \ln \left[\epsilon_B + (\epsilon_A - \epsilon_B)\phi_A\right]\label{eq:free_homo}
\end{eqnarray}
where the subscript $H$ stands for homogeneous. Here, the first term is the self-energy of the monomers taking into account
the neutral interactions, the second term is the well-known Flory-Huggins interaction
energy and the next two terms represent the translational entropy of the chains. The fifth term
is a constant obtained from the integration over the orientation of dipoles. The last term originates from the dipolar interactions. 

If the dielectric constant of the blend is small i.e., $\epsilon_B + (\epsilon_A - \epsilon_B)\phi_A \ll 1$, we can expand 
the last term in Eq. ~\ref{eq:free_homo} in powers of 
$\phi_A$ and second degree terms in $\phi_A$ appearing in Eq. ~\ref{eq:free_homo} have 
the same functional forms as in Eq. ~\ref{eq:free_dipoles}. 
The condition of $\epsilon_B + (\epsilon_A - \epsilon_B)\phi_A \ll 1$  is 
met at high temperature or for 
monomers with very small dipole moments. This comparison reveals that analysis based on freely 
rotating dipoles has a very narrow parameter space where it is valid. Furthermore, the comparison 
reveals that $q_{max}^2/\Delta z = 1/\lambda^3$ in order for the two expressions to be the same.  

\begin{figure}[ht]  \centering
\subfigure[]{\includegraphics[height=2.75in]{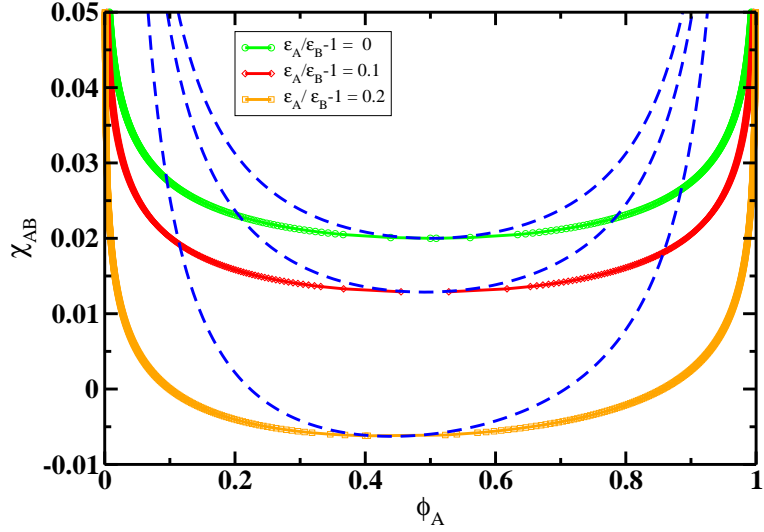}} \vspace{0.3in}\\
\subfigure[]{\includegraphics[height=2.75in]{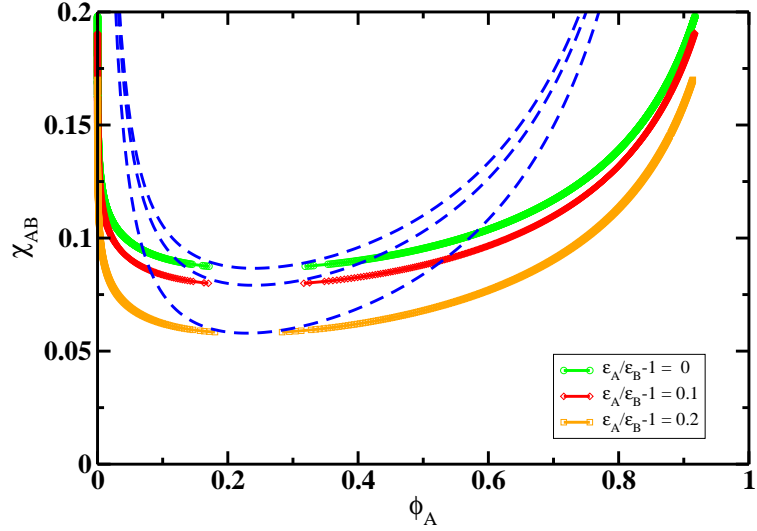}}
\caption{ Coexistence curves along with spinodals for symmetric ($N_A = N_B = 100$) and asymmetric chain
lengths ($N_A =  100, N_B = 10$). The case of $\epsilon_A/\epsilon_B = 1$ corresponds to the Flory-Huggins theory.}
\label{fig:phase_diagram}
\end{figure}
The \textit{exchange} chemical potential can be computed using the free energy given by Eq. ~\ref{eq:free_homo}. 
It is given by 
\begin{eqnarray}
\mu = \frac{\partial}{\partial \phi_A} \left[\frac{F_H}{\rho_o V k_BT}\right] &=& \frac{1}{2}\left[w_{AA} - w_{BB} 
\right]\rho_o + \chi_{AB}(1-2\phi_A) 
+ \frac{1}{N_A}\ln \left(\phi_A\right) -\frac{1}{N_B}\ln \left(1-\phi_A\right) \nonumber \\
&&+  \frac{q_{max}^2 \Delta \epsilon }{8\pi \Delta z \rho_o \left(1 + \Delta \epsilon \phi_A\right)} \label{eq:mu_homo}
\end{eqnarray}
Spinodal can also be computed and is given by
\begin{eqnarray}
\frac{\partial^2}{\partial \phi_A^2} \left[\frac{F_H}{\rho_o V k_BT}\right] &=&  
\frac{1}{N_A \phi_A} + \frac{1}{N_B \left(1-\phi_A\right)} -2\chi_{AB} 
-  \frac{q_{max}^2 \Delta \epsilon^2}{8\pi \Delta z \rho_o \left(1 + \Delta \epsilon \phi_A\right)^2} = 0\label{eq:spino_homo}
\end{eqnarray}
It is clear from these equations that it is the parameter $\Delta \epsilon = \epsilon_A/\epsilon_B -1$, which captures the 
effect of dipolar interactions on phase equilibria. Also, note that the same parameter appears in the free energy of mixing ($\Delta F_{mix}$) identified from Eq. ~\ref{eq:free_homo} by writing it in the form
\begin{eqnarray}
      \frac{F_H}{\rho_0 V k_BT} &=& \phi_A \left[\frac{w_{AA}\rho_0}{2} -\frac{1}{N_A} - \ln 4\pi + \frac{q_{max}^2}{8\pi \Delta z \rho_o} \ln \epsilon_A \right] \nonumber \\
&& + (1-\phi_A) \left[\frac{w_{BB}\rho_0}{2} - \frac{1}{N_B} - \ln 4\pi + \frac{q_{max}^2}{8\pi \Delta z \rho_o} \ln \epsilon_B \right]  + \frac{\Delta F_{mix}}{k_B T}
\end{eqnarray}
where
\begin{eqnarray}
\frac{\Delta F_{mix}}{k_B T} &=& \chi_{AB}\phi_A (1-\phi_A) 
+ \frac{\phi_A}{N_A} \ln \left(\frac{\phi_A}{N_A}\right) + \frac{1-\phi_A}{N_B} \ln \left(\frac{1-\phi_A}{N_B}\right)
\nonumber \\
&& +  \frac{q_{max}^2}{8\pi \Delta z \rho_o} \left\{\ln \left[1 + \Delta \epsilon\phi_A\right]
 - \phi_A \ln \left[1 + \Delta\epsilon\right]\right\}\label{eq:free_mix}
\end{eqnarray}

\begin{figure}[ht]  \centering
\subfigure[]{\includegraphics[height=2.75in]{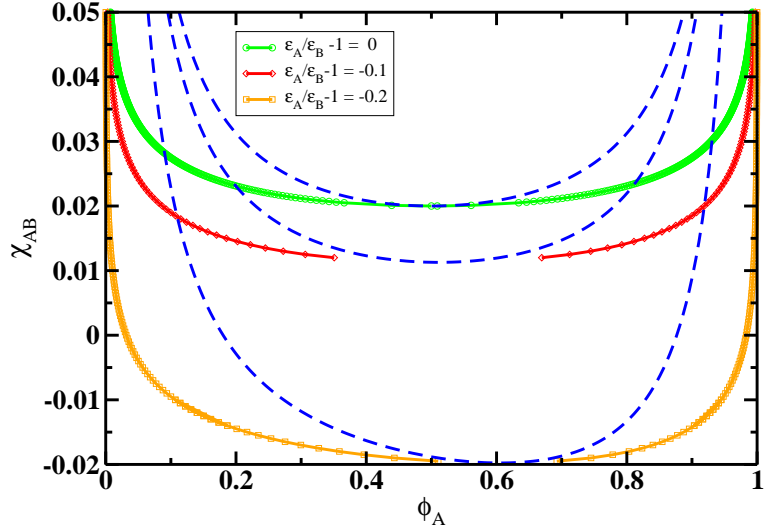}} \vspace{0.3in}\\
\subfigure[]{\includegraphics[height=2.75in]{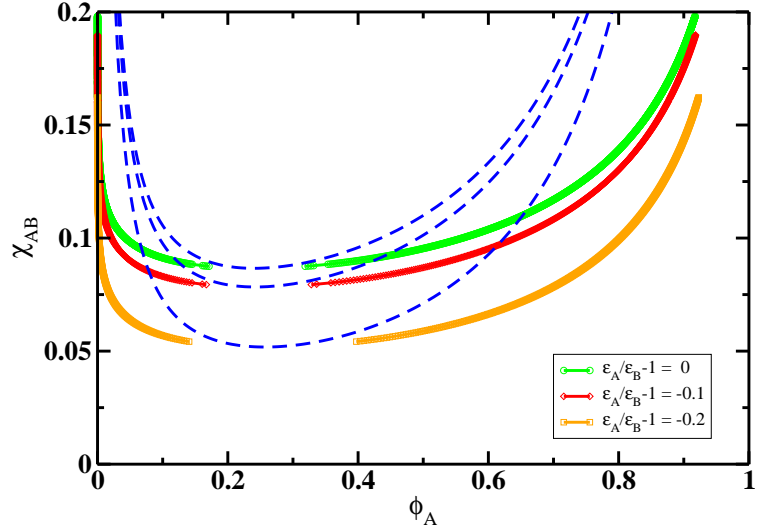}}
\caption{ Dependence of phase diagram on the sign of the dielectric mismatch is shown here by computing
coexistence curves along with spinodals for symmetric ($N_A = N_B = 100$) and asymmetric chain
lengths ($N_A =  100, N_B = 10$).}
\label{fig:phase_diagram_ganeg}
\end{figure}

\begin{figure}[ht]  \centering
\vspace{0.3in}
\subfigure[]{\includegraphics[height=2.75in]{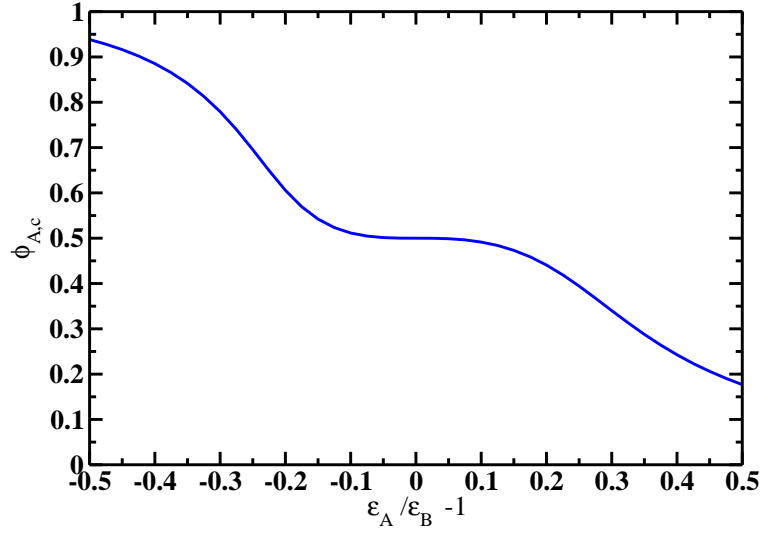}} \vspace{0.3in}\\
\subfigure[]{\includegraphics[height=2.75in]{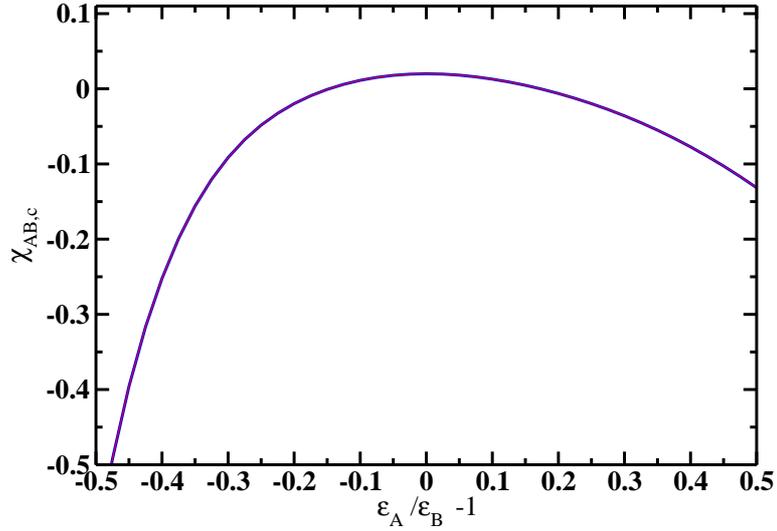}}
\caption{ Shifting of the critical point due to the dielectric mismatch is shown for 
systems with symmetric chain lengths ($N_A = N_B = 100$).  The critical point is characterized 
by volume fraction $\phi_{A,c}$ and $\chi_{AB} = \chi_{AB,c}$ at the minimum of the spinodal curve, shown in 
(a) and (b), respectively.}
\label{fig:critical_point_symm}
\end{figure}

Coexistence curves along with the spinodals are computed using above equations and presented in Figures 
~\ref{fig:phase_diagram} and ~\ref{fig:phase_diagram_ganeg}. For the computations, we have taken 
$q_{max} =  2\pi/b$ and $\Delta z = b = \rho_o^{1/3}$. In these Figures, case of $\epsilon_A/\epsilon_B = 1$ corresponds to the Flory-Huggins theory as evident from the free energy of mixing (cf. Eq. ~\ref{eq:free_mix}). 
From the phase diagrams computed for 
symmetric and asymmetric chain lengths in Fig. ~\ref{fig:phase_diagram}, it is clear that differences in 
dielectric constants between the 
two polymers tend to induce phase segregation and regime where uniformly mixed homogeneous phase is stable, shrinks with an increase in the dielectric mismatch. Note that the limiting case where the degree of polymerization of one of the 
polymers is unity corresponds to polymer solutions. Keeping this in mind, the theory presented here can be used to 
infer solubility of a polymer in different solvents\cite{comment1}. It is clear from the results that polymers with dielectric constants similar to solvent have the largest uniformly mixed phase region. This is in agreement with the 
empirical rule of ``like dissolves like''. Furthermore, comparison between Figures
 ~\ref{fig:phase_diagram} and ~\ref{fig:phase_diagram_ganeg} reveals that the sign of the dielectric mismatch also plays a role in shifting the phase boundaries. Effects of the sign of the the dielectric mismatch is clearer if one looks at the volume fraction and $\chi_{AB}$ at the critical point as shown in Fig. ~\ref{fig:critical_point_symm} for 
symmetric chain lengths $N_A = N_B = 100$. The critical 
point corresponds to minimum in the spinodal curve. From Fig. ~\ref{fig:critical_point_symm}(a), it is found that 
the volume fraction of component $A$ at the critical point decreases monotonically with an increase in $\Delta \epsilon$. Also, the $\chi_{AB}$ at the critical point increases and then decreases with an increase in $\Delta \epsilon$, with 
the maximum corresponding to $\Delta \epsilon = 0$ (cf. Fig. ~\ref{fig:critical_point_symm}(b)). 
These shifts in the critical point originate from the dependence of the exchange chemical potential 
(cf. Eq. ~\ref{eq:mu_homo}) on $\Delta \epsilon$ and $\phi_{A}$. The maximum in 
Fig. ~\ref{fig:critical_point_symm}(b) highlights the fact that the dielectric mismatch leads to enhanced phase 
segregation even with negative $\chi_{AB}$. Similar shifts are found for systems with asymmetric chain lengths.   

Note that the prediction of an enhanced phase segregation with an increase in the dielectic mismatch is in agreement with the 
treatment of freely rotating dipoles. However, note that the critical point also gets shifted (cf. Fig. ~\ref{fig:critical_point_symm}) due to the mismatch, 
in contrast to being fixed for freely rotating dipoles. Due to the fact that the field theoretical formalism has 
the freely rotating dipoles as a limiting case, such a comparison highlights the invalidity of the assumption 
that dipoles can rotate freely in the two phase region. In the following, we use the field theoretical formalism to 
compute interfacial properties of a planar interface between the co-existing phases in the strong segregation limit. 

\subsection{Planar interface between the co-existing phases in the weak-coupling limit}
A planar interface between the coexisting phases can be studied using formalism developed by 
Helfand-Tagami-Sapse\cite{helfand_tagami1,helfand_tagami2,helfand_tagami3,
helfand_sapse} for very long polymers ($N_A = N_B \rightarrow \infty$). Using the 
formalism, we have studied interfacial tension and density profiles in the strong segregation limit 
so that far from the interface, there are pure phases with corresponding volume fraction of a component 
approaching unity. In this limit, interfacial tension is independent of the value of $N_A$ and $N_B$.

A straightforward mapping on to the formalism reveals that 
density profile of component A is given by 
\begin{eqnarray}
      z &=& \frac{1}{2}\int_{1/2}^{\phi_A(z)}d\phi \left\{\frac{\beta_A^2}{\phi} 
+ \frac{\beta_B^2}{1-\phi} + \frac{q_{max}\Delta \epsilon^2}{8\pi \left[1 + \Delta \epsilon \phi\right]^2}\right\}^{1/2}
        \left\{\frac{\Delta f^\star\left\{\phi\right\}}{k_BT} \right\}^{-1/2}
\end{eqnarray}
with the zero of $z$ taken so that $\phi_A(0) =  1/2$ and $\beta_k^2 =  \rho_o b_k^2/6$ for $k=A,B$ is 
conformational asymmetry parameter\cite{helfand_sapse}. Also, $\Delta f^\star$ is the integrand 
for the free energy density of mixing in an inhomogeneous medium for $N_A \rightarrow \infty, N_B \rightarrow \infty$, 
given by
\begin{eqnarray}
      \frac{\Delta f^\star\left\{\phi_A\right\}}{\rho_o k_BT} &=& \chi_{AB}\phi_A(z) (1-\phi_A(z)) + \frac{q_{max}^2}{8\pi \Delta z \rho_o} \left\{\ln \left[1 + \Delta \epsilon \phi_A(z)\right] - \phi_A(z) \ln \left[1 + \Delta \epsilon\right]\right\}
\end{eqnarray}
Interfacial tension for the interface between coexisting phases is given by 
\begin{eqnarray}
      \frac{\gamma}{k_B T} &=& \int_{0}^{1}d\phi \left\{\frac{\beta_A^2}{\phi} + \frac{\beta_B^2}{1-\phi} 
+ \frac{q_{max}\Delta \epsilon^2}{8\pi \left[1 + \Delta \epsilon \phi\right]^2}\right\}^{1/2}
        \left\{\frac{\Delta f^\star\left\{\phi\right\}}{k_BT} \right\}^{1/2}
\end{eqnarray}

\begin{figure}[ht]  \centering
\vspace{0.4in}
$\begin{array}{c@{\hspace{0.3in}}c@{\hspace{0.3in}}}                                         
\includegraphics[width=5in,height=4in]{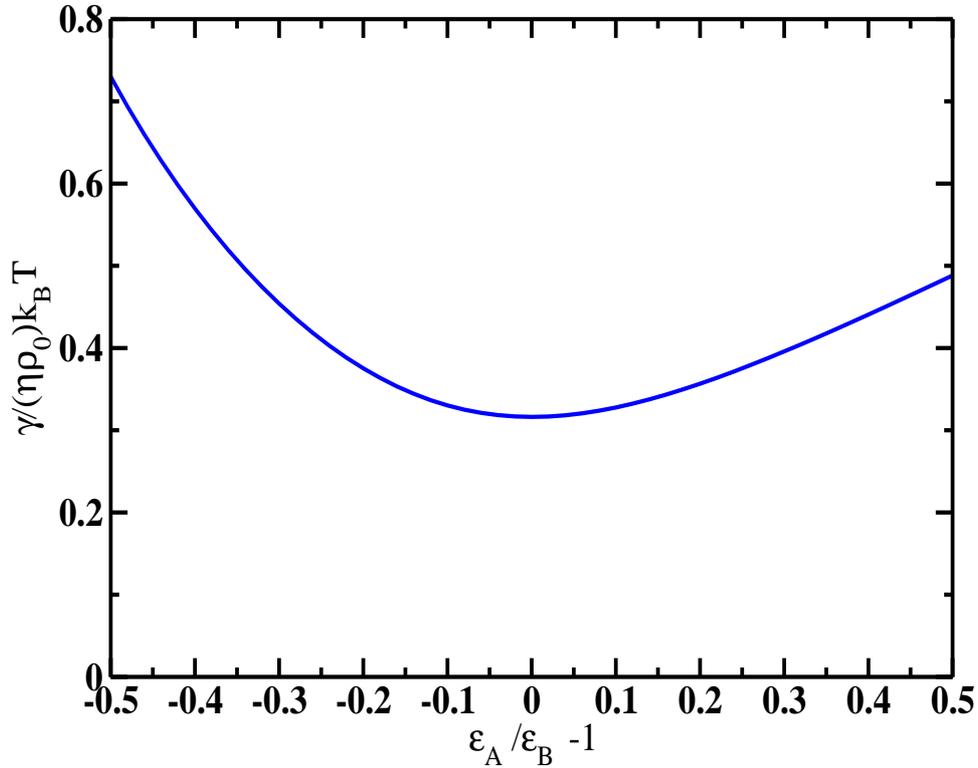}                                                   
\end{array}$
\caption{Dependence of the interfacial tension on the dielectric mismatch for the conformationally symmetric polymer 
blends is shown here. $\chi_{AB} =  0.1$ is used
in these calculations.}
\label{fig:interfacial_tension}
\end{figure}

\begin{figure}[ht]  \centering
\vspace{0.6in}
$\begin{array}{c@{\hspace{0.3in}}c@{\hspace{0.3in}}}                                         
\includegraphics[width=5in,height=4in]{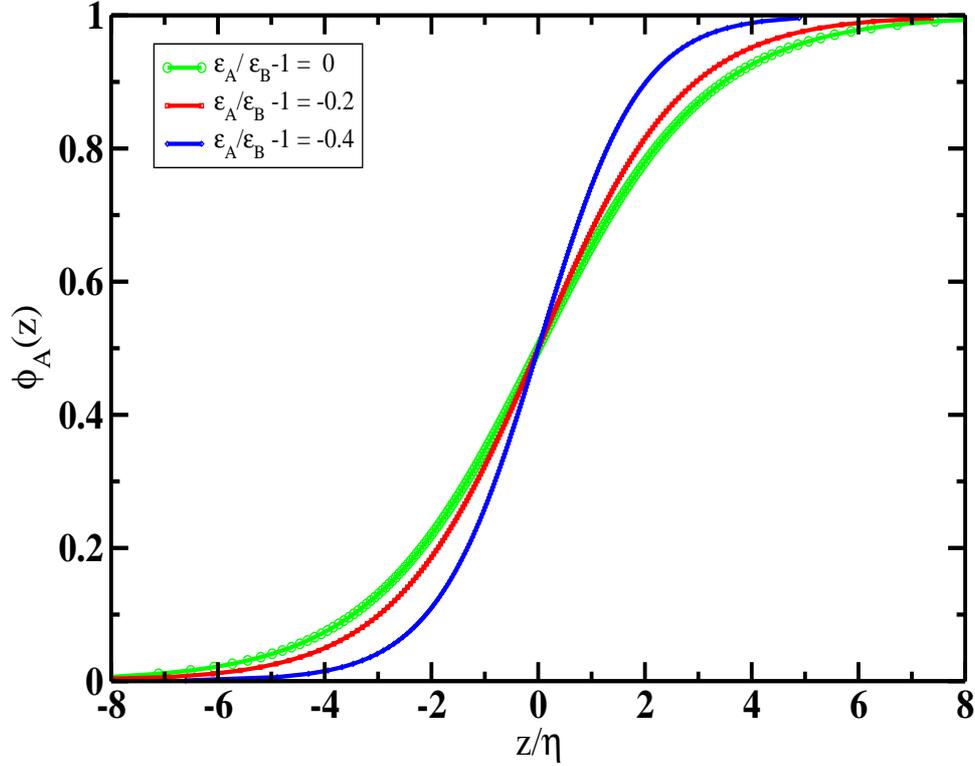}                                                   
\end{array}$
\caption{ Dependence of the volume fraction profiles on the dielectric mismatch is highlighted here. It is shown that 
an increase in the mismatch leads to sharper polymeric interfaces. $\chi_{AB} =  0.1$ is used
in these calculations.}
\label{fig:density_profiles}
\end{figure}

In order to highlight the effects of dielectric mismatch on interfacial properties, we consider a conformaltionally 
symmetric (i.e., $\beta_A = \beta_B = \beta$) polymer blend in this work. For the blend, we have computed density profiles and interfacial tension in units of parameter $\eta = \beta/\sqrt{\rho_0}$. 
Results for the interfacial tension are presented in Fig. ~\ref{fig:interfacial_tension}. It is clear that 
the interfacial tension increases with an increase in the dielectric mismatch. However, amount of an increase in 
the interfacial tension depends on the sign of the dielectric mismatch parameter ($\Delta \epsilon$) 
due to asymmetric nature of the curve in Fig. ~\ref{fig:interfacial_tension} about $\Delta \epsilon = 0$.
The asymmetric nature arises due to the asymmetry of function $\Delta f^{\star}$ about $\Delta \epsilon = 0$. The increase in the interfacial tension with an increase in the mismatch parameter
is a direct outcome of an increase in $\Delta f^{\star}$, which also leads to sharper interfacial density 
profiles as shown in Fig. ~\ref{fig:density_profiles}. 

As the dielectric mismatch originates from a difference in dipole moments, it is clear that 
dipolar interactions induce phase segregation that subsequently causes sharper polymeric interfaces. The latter effect 
of an increase in sharpness of interfacial density profiles with an increase in the mismatch is in disagreement with 
the Cahn-Hilliard like treatment of freely rotating dipoles near the critical point, as presented in this paper. This comparison reveals another drawback of the treatment based on freely rotating dipoles for an inhomogeneous medium. Near the critical point, in addition to the assumption of freely-rotating dipoles in the two phase region, one needs to go beyond the saddle-point approximation invoked in the computation of the interfacial tension. 
For example, it is well-known\cite{widom_book} that one needs to take into account fluctuations of the order parameter profiles in order to get the experimentally known 
exponents. 

\section{Conclusions} \label{sec:conclusions}
A new method is developed to compute electrostatic contribution to the free energy. 
The method maps the evaluation of functional integrals over electrostatic collective variables on 
to the calculations of ratio of functional determinants frequently 
encountered in quantum mechanical systems\cite{kleinertbook}. Furthermore, the ratio of 
functional determinants is computed by mapping it on to the solution of a boundary value problem instead 
of computing infinite eigen values of different operators. 
The method allows the computation of 
free energy for a \textit{given} dielectric function profile and can be readily applied to a number of problems 
where such a profile is known. 

We have applied this method to study the effects of dipolar interactions in
polymer blends. Parameters $l_{Bo}p_A^2,l_{Bo}p_B^2$ and $l_{Bo}p_A p_B$ control the strengths of 
correlations due to dipolar interactions (as evident from Eq. ~\ref{eq:segelecideal}) and local ordering of dipoles is 
expected\cite{MM_dipoles} to 
compete with the macrophase separation in the low temperature regime. In this work, we have studied 
the high-temperature regime, where correlations due to 
orientation dependent dipolar interactions are negligible (weak coupling limit).
In the weak coupling limit for the dipoles, the linear mixing rule based on the 
volume fractions serves as a
good approximation for the local dielectric function. Using the linear mixing rule and newly developed method 
for the computation of electrostatic contributions, we have constructed a free energy expression for
homogeneous phase and studied phase coexistence. Furthermore, a planar interface between the coexisting phases is
studied by generalizing Helfand-Tagami-Sapse's formalism to dipolar polymer blends. It is found that dipolar interaction
sinduces phase segregation.  Furthermore, an increase in mismatch in monomeric dipole moments in polymer blends leads to
an increase in interfacial tension with sharper interfaces between the co-existing phases. Magnitude of the effects of dipolar mismatch depends on the choice of cutoff (in the form of $q_{max}$ in this work) below which short range interactions dominate over the dipole-dipole interactions. Detailed comparison with experiments in future can reveal sensitivity of the coexistence curves to 
the choice of cutoff. 

Comparison of these theoretical predictions with those obtained from the treatment of freely rotating 
dipoles reveals that 
the latter is valid in the \textit{homogeneous} regime, where $\epsilon_A,\epsilon_B \sim 1$ (which corresponds to high temperature 
and weak dipole moments of monomers). Theoretical treatment based on the assumption of freely rotating dipoles in an 
an inhomogeneous medium leads to erroneous predictions such as the insensitivity of critical point to the mismatch in 
monomeric dipole moments and a decrease in sharpness of interface between co-existing phases with an increase in the mismatch. Therefore, predictions from the field theory should be used for studies 
involving phase co-existence and 
a planar interface between the co-existing phases. 
  
It is to be noted that the non-local contribution to the
free energy in Eq. ~\ref{eq:final_nonlocal} is of the same functional form as in Ref. \cite{podgornik}.
In Ref. \cite{podgornik}, it was shown that the non-local term can be derived by considering non-retarded
limit in a generalization of
Lifshitz's theory\cite{intermolecular_forces} of van der Waals interactions, originally derived for discrete dielectric slabs, to
study systems with continuous variation in the dielectric function. However, local incompressiblity constraint used in this work differs from the classical theory of van der Waals.
Also, we have ignored the effects of finite polarizability or the ``induced'' dipole moments, 
which has been shown to enhance 
the net dipole moments\cite{onsager_moments}. Neglecting the induction effect is a reasonable first step before the development of 
a comprehensive theory of the dielectric response of flexible macromolecules. In future, we plan to overcome this limitation of the theory using the field theoretical method developed here.

\section*{ACKNOWLEDGMENTS}
\setcounter {equation} {0} \label{acknowledgement}
This research used resources of the Oak Ridge Leadership Computing
Facility at the Oak Ridge National Laboratory (ORNL), which is supported by the
Office of Science of the U.S. Department of Energy under Contract No.
DE-AC05-00OR22725. R.K. and B.G.S. acknowledge support from the Center for Nanophase 
Materials Sciences, which is sponsored at Oak Ridge National Laboratory by the Scientific User Facilities Division, office of Basic Energy Sciences, U.S. Department of Energy (DOE). M.M acknowledges support from the National Science Foundation (Grant No. DMR 1105362). We acknowledge support for aspects of code
development from the ORNL Laboratory Directed Research and Development.

\renewcommand{\theequation}{A-\arabic{equation}}
  \setcounter{equation}{0}  
\renewcommand{\theequation}{A-\arabic{equation}}
  \setcounter{equation}{0}  
  \section*{APPENDIX A : Field theory for dipolar polymer blends} \label{app:A}
A field theory for the polymer blends can be constructed following a standard 
protocol\cite{fredbook}. We start from the electrostatics contributions to the partition function. 
For the electrostatics contribution to the partition function written in the form
Eq. ~\ref{eq:final_particle_elec}, we use Hubbard-Stratonovich transformation in the form
\begin{eqnarray}
\exp\left[-H_{dd}\right] &=& \frac{1}{\mu_\psi} \int D\left[\psi\right]\exp \left[i\int d\mathbf{r}
\left\{\nabla_{\mathbf{r}}.\hat{P}_{ave}(\mathbf{r})\right\} \psi(\mathbf{r}) +
\frac{1}{8\pi l_{Bo}}\int d\mathbf{r}\psi(\mathbf{r})\nabla_\mathbf{r}^2 \psi(\mathbf{r}))\right] \nonumber \\
&&
\end{eqnarray}
where
\begin{eqnarray}
\mu_\psi &=& \int D\left[\psi\right]\exp \left[\frac{1}{8\pi
l_{Bo}}\int d\mathbf{r}\psi(\mathbf{r})\nabla_\mathbf{r}^2 \psi(\mathbf{r}))\right]
\end{eqnarray}
Using this transformation, we can evaluate the integrals over the orientations of the dipoles exactly. 
In particular, using 
\begin{eqnarray}
\int \prod_{j=A,B} \prod_{\alpha=1}^{n_j}\prod_{s_{j,\alpha}=0}^{N_j} d\mathbf{u}_{j,\alpha}(s_{j,\alpha})
\exp\left[i\int d\mathbf{r}
\left\{\nabla_{\mathbf{r}}.\hat{P}_{ave}(\mathbf{r})\right\} \psi(\mathbf{r})\right] && \nonumber \\
 = \int \prod_{j=A,B} \prod_{\alpha=1}^{n_j}\prod_{s_{j,\alpha}=0}^{N_j} \left\{d\mathbf{u}_{j,\alpha}(s_{j,\alpha})
\exp\left[ -i p_j\mathbf{u}_{j,\alpha}(s_{j,\alpha}).\nabla_{\mathbf{r}}\psi(\mathbf{r})\right] \right\} && \nonumber \\ 
= \exp\left[\int d\mathbf{r} \sum_{j=A,B}\hat{\rho}_j(\mathbf{r}) \ln \left[\frac{4\pi \sin\left(p_j |\nabla_\mathbf{r}\psi(\mathbf{r})|\right)}
{p_j |\nabla_\mathbf{r}\psi(\mathbf{r})|}\right]\right],&& 
%
\end{eqnarray} 
the partition function given by Eq. ~\ref{eq:parti_melts} becomes
\begin{eqnarray}
       Z & = & \frac {1}{n_A!n_B!}\int \prod_{j=A,B}\prod_{\alpha=1}^{n_j} D[\mathbf{R}_{j,\alpha}] 
\frac{1}{\mu_\psi} \int D\left[\psi\right]\exp \left [-H_0\left\{\mathbf{R}_{j,\alpha}\right\} 
- H_w\left\{\mathbf{R}_{j,\alpha},\mathbf{R}_{j',\alpha'}\right \} \right .\nonumber \\
&&  +
\frac{1}{8\pi l_{Bo}}\int d\mathbf{r}\psi(\mathbf{r})\nabla_\mathbf{r}^2 \psi(\mathbf{r})) 
+ \int d\mathbf{r} \hat{\rho}_A(\mathbf{r}) \ln \left[\frac{4\pi \sin\left(p_A |\nabla_\mathbf{r}\psi(\mathbf{r})|\right)}
{p_A |\nabla_\mathbf{r}\psi(\mathbf{r})|}\right]\nonumber \\
&& \left . + \int d\mathbf{r} \hat{\rho}_B(\mathbf{r}) \ln \left[\frac{4\pi \sin\left(p_B |\nabla_\mathbf{r}\psi(\mathbf{r})|\right)}
{p_B |\nabla_\mathbf{r}\psi(\mathbf{r})|}\right]\right ] \prod_{\mathbf{r}}\mathbf{\delta}
\left[\hat{\rho}_{A}(\mathbf{r}) + \hat{\rho}_{B}(\mathbf{r}) - \rho_0\right]\label{eq:parti_elec_added}
\end{eqnarray}
Rewriting $H_w$ given by Eq. ~\ref{eq:dispersion} using the local incompressibility 
constraint as
\begin{eqnarray}
H_w\left\{\mathbf{R}_{j,\alpha},\mathbf{R}_{j,\alpha'}\right\} &=& \frac{1}{2}\left[w_{AA}n_A N_A + w_{BB}n_B N_B\right]\rho_0 + \chi_{AB}b^3\int d\mathbf{r} \hat{\rho}_A(\mathbf{r})\hat{\rho}_{B}(\mathbf{r})\label{eq:dispersion_2}
\end{eqnarray}
where we have defined a dimensionless chi parameter by
\begin{eqnarray}\chi_{AB}b^3 &=& w_{AB} - \frac{w_{AA} + w_{BB}}{2} \label{eq:chi_parameter}
\end{eqnarray}
Now, we introduce collective variables corresponding to $\hat{\rho_j}$ by using the identity
\begin{eqnarray}
Z\left\{\hat{\rho}_A,\hat{\rho}_B\right\} &=& \int D\left[\rho_A\right]\int D\left[\rho_B\right] 
Z\left\{\rho_A,\rho_B\right\}\prod_{\mathbf{r}}\mathbf{\delta}\left[\rho_A(\mathbf{r}) - \hat{\rho}_A(\mathbf{r})\right]\prod_{\mathbf{r}}\mathbf{\delta}\left[\rho_B(\mathbf{r}) - \hat{\rho}_B(\mathbf{r})\right] \label{eq:parti_2} \nonumber
\end{eqnarray}
and write the local constraints (represented by delta functions) in terms of functional integrals using
\begin{eqnarray}
\prod_{\mathbf{r}}\mathbf{\delta}\left[\rho_j(\mathbf{r}) - \hat{\rho_j}(\mathbf{r})\right] 
 &=& \int D\left[w_j\right] \exp\left[i \int d\mathbf{r} w_j(\mathbf{r})\left\{\rho_j(\mathbf{r})- \hat{\rho_j}(\mathbf{r})\right\}\right] \label{eq:order_field}
\end{eqnarray}
for $j = A,B$.

Using these transformations and evaluating a trivial functional intergral over $\rho_B$, 
we can write the partition function as

\begin{eqnarray}
       Z & = & \frac{1}{\mu_\psi}\int D\left[\rho_A\right]\int  D\left[w_A\right]\int D\left[w_B\right]\int  D\left[\psi\right] 
\exp \left [-\frac{F_0}{k_BT} - \frac{H}{k_BT}\right ] \label{eq:parti_abstract}
\end{eqnarray}
where
\begin{eqnarray}
      \frac{F_0}{k_BT} &=& \frac{1}{2}\left[w_{AA} n_A N_A + w_{BB}n_B N_B\right]\rho_0  + 
\ln \left[\frac{n_A!n_B!}{(4\pi)^{\sum_{j=A,B}n_j N_j}\prod_{j=A,B}Q_{j}^{n_j}\left\{0\right\}}\right]
\end{eqnarray}
and
\begin{eqnarray}
       \frac{H}{k_BT} &=& \chi_{AB}b^3\int d\mathbf{r} \rho_A(\mathbf{r})(\rho_0 - \rho_A(\mathbf{r})) - i \int d\mathbf{r} w_A(\mathbf{r})\rho_A(\mathbf{r}) 
- i \int d\mathbf{r} w_B (\mathbf{r})\left[\rho_0 - \rho_A(\mathbf{r}) \right] 
\nonumber \\
&& - \frac{1}{8\pi l_{Bo}}\int d\mathbf{r} \psi(\mathbf{r})\nabla_{\mathbf{r}}^2 \psi(\mathbf{r}) - \int d\mathbf{r}\rho_A(\mathbf{r})\ln \left[\frac{\sin\left(p_A |\nabla_\mathbf{r}\psi(\mathbf{r})|\right)}
{p_A |\nabla_\mathbf{r}\psi(\mathbf{r})|}\right] 
\nonumber \\
&& - \int d\mathbf{r} \left\{\rho_0 - \rho_A(\mathbf{r})\right\} 
\ln \left[\frac{\sin\left(p_B |\nabla_\mathbf{r}\psi(\mathbf{r})|\right)}
{p_B |\nabla_\mathbf{r}\psi(\mathbf{r})|}\right]  - \sum_{j=A,B}n_j\ln \bar{Q}_{j}\left\{w_j\right\}
\end{eqnarray}
where $Q_{j}\left\{0\right\} = \int D\left[\mathbf{R}\right] \exp\left[-\frac{3}{2b_j^2}\int_{0}^{N_j} ds 
\left(\frac{\partial \mathbf{R}}{\partial s}\right)^2\right] = V$, $V$ being the total volume and  
\begin{eqnarray}
      \bar{Q}_{j}\left\{w_j\right\} = \frac{Q_{j}\left\{w_j\right\}}{Q_{j}\left\{0\right\}} &=& \frac{\int D\left[\mathbf{R}\right] \exp\left[-\frac{3}{2b_j^2}\int_{0}^{N_j} ds 
\left(\frac{\partial \mathbf{R}}{\partial s}\right)^2 - i \int_{0}^{N_j} ds w_j(\mathbf{R})\right]
}{\int D\left[\mathbf{R}\right] \exp\left[-\frac{3}{2b_j^2}\int_{0}^{N_j} ds
\left(\frac{\partial \mathbf{R}}{\partial s}\right)^2\right]} 
\end{eqnarray}
 
\section*{REFERENCES}
\setcounter {equation} {0}
\pagestyle{empty} \label{REFERENCES}

\newpage
\section*{For Table of Contents Only}
\begin{center}
\textbf{Enhanced phase segregation induced by dipolar interactions in polymer blends}\\
\textbf{Rajeev Kumar, Bobby G. Sumpter and M. Muthukumar}
\end{center}

\begin{figure}[ht]  \centering
$\begin{array}{c@{\hspace{0.3in}}c@{\hspace{0.3in}}}                                         
\includegraphics[height=1.75in,width=3.75in]{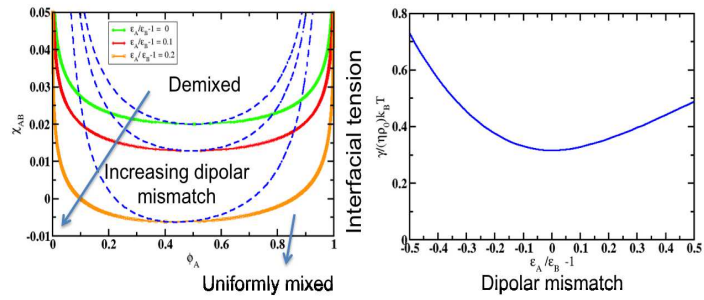}                                                   
\end{array}$
\caption{Table of Contents Graphic}
\end{figure}


\begin{thebibliography}{99}
\bibitem{comment1}
Note that effects of density fluctuations relevant for phase segregation in polymer solutions 
are not taken into account with in the saddle-point approximation used here.
\end{thebibliography}
\end{document}